\documentclass[notitlepage,nofootinbib,preprintnumbers,amssymb,superscriptaddress,twocolumn]{revtex4-2}
\usepackage{amsfonts,amssymb,mathtools,graphicx,xcolor,bm}
\definecolor{ultramarine}{rgb}{0.07, 0.04, 0.56}
\definecolor{cadmiumgreen}{rgb}{0.0, 0.42, 0.24}
\definecolor{indigo(dye)}{rgb}{0.0, 0.25, 0.42}
\usepackage[linktocpage=true,breaklinks]{hyperref}
\hypersetup{
colorlinks=true,
citecolor=ultramarine,
linkcolor=cadmiumgreen,
urlcolor=indigo(dye),
}

\usepackage{autobreak}
\usepackage{tikz}
\usepackage{here}

\begin{document}

\preprint{YITP-22-25, IPMU22-0014}

\title{Built-in scordatura in U-DHOST}

\author{Antonio De Felice}
\affiliation{Center for Gravitational Physics and Quantum Information, Yukawa Institute for Theoretical Physics, Kyoto University, 606-8502, Kyoto, Japan}

\author{Shinji Mukohyama}
\affiliation{Center for Gravitational Physics and Quantum Information, Yukawa Institute for Theoretical Physics, Kyoto University, 606-8502, Kyoto, Japan}
\affiliation{Kavli Institute for the Physics and Mathematics of the Universe (WPI), The University of Tokyo, 277-8583, Chiba, Japan}

\author{Kazufumi Takahashi}
\affiliation{Center for Gravitational Physics and Quantum Information, Yukawa Institute for Theoretical Physics, Kyoto University, 606-8502, Kyoto, Japan}

\begin{abstract}
Modified gravity theories can accommodate exact solutions, for which the metric has the same form as the one in general relativity, i.e., stealth solutions.
One problem with these stealth solutions is that perturbations around them exhibit strong coupling when the solutions are realized in degenerate higher-order scalar-tensor theories.
We show that the strong coupling problem can be circumvented in the framework of the so-called U-DHOST theories, in which the degeneracy is partially broken in such a way that higher-derivative terms are degenerate only in the unitary gauge.
In this sense, the scordatura effect is built-in in U-DHOST theories in general.
There is an apparent Ostrogradsky mode in U-DHOST theories, but it does not propagate as it satisfies a three-dimensional elliptic differential equation on a spacelike hypersurface.
We also clarify how this nonpropagating mode, i.e., the ``shadowy'' mode shows up at the nonlinear level.
\end{abstract}

\maketitle


{\it Introduction}.---The direct detection of gravitational waves from compact binary coalescences~\cite{LIGOScientific:2018mvr,LIGOScientific:2020ibl,LIGOScientific:2021usb,LIGOScientific:2021djp} and the direct imaging of black hole shadow~\cite{Akiyama:2019cqa} offer a possibility for testing gravity at strong-field regimes.
This motivates us to investigate alternative theories of gravity and predict their observational consequences for comparison with general relativity (GR).
The observations so far are consistent with the Kerr metric, which is a unique asymptotically flat and stationary black hole solution in GR.
This fact, of course, adds new supporting evidence for the validity of GR. 
However, it does not exclude all modified gravity theories as they can allow the metric of the same form as the GR solutions as an exact solution in principle.
To be more precise, the solution can have nontrivial hair associated with additional degrees of freedom (DOFs) in modified gravity, but still the background metric itself remains the same as in GR.
In this sense, effects of modified gravity are invisible in the background metric, and hence such a configuration is ``stealth.''
Nevertheless, nontrivial effects of modified gravity would show up once gravitational perturbations are taken into account, by which stealth solutions can be distinguished from GR solutions.

In the context of scalar-tensor theories, the first stealth solution called ghost condensation was constructed on Minkowski and de Sitter backgrounds~\cite{ArkaniHamed:2003uy,Arkani-Hamed:2003juy} and then extended to a Schwarzschild black hole background~\cite{Mukohyama:2005rw}.
The general construction of stealth solutions was then developed~\cite{Motohashi:2018wdq,Takahashi:2020hso} and perturbations around them~\cite{deRham:2019gha,Khoury:2020aya,Takahashi:2021bml} have been extensively studied, with a particular focus on degenerate higher-order scalar-tensor (DHOST) theories~\cite{Horndeski:1974wa,Deffayet:2011gz,Kobayashi:2011nu,Langlois:2015cwa,Crisostomi:2016czh,BenAchour:2016fzp,Takahashi:2017pje,Langlois:2018jdg,Takahashi:2021ttd}, which form a large class of healthy scalar-tensor theories without Ostrogradsky ghost~\cite{Woodard:2015zca} thanks to degeneracy conditions imposed on higher-derivative terms~\cite{Motohashi:2014opa,Langlois:2015cwa,Motohashi:2016ftl,Klein:2016aiq,Motohashi:2017eya,Motohashi:2018pxg}.
The framework of DHOST theories is now so common that various aspects including cosmology and black holes have been extensively studied in the literature (see \cite{Langlois:2018dxi,Kobayashi:2019hrl} and references therein).
It should also be noted that stealth solutions are often realized as attractors of the system~\cite{ArkaniHamed:2003uy}.
A remarkable fact is that perturbations about a stealth Minkowski (or de Sitter) background in DHOST theories are strongly coupled in general, meaning that the perturbative description is no longer valid and that one cannot compute anything reliably based on the effective field theory~(EFT)~\cite{ArkaniHamed:2003uy,Motohashi:2019ymr}.
The point is that the strong coupling scale is proportional to some positive power of the sound speed of the scalar perturbation, which is typically suppressed by the Planck scale in DHOST theories.
The strong coupling problem persists in stealth black hole solutions~\cite{deRham:2019gha,Khoury:2020aya,Takahashi:2021bml}, for which the metric is asymptotically Minkowski (or de Sitter).
Therefore, in order to avoid the generic problem of strong coupling within the framework of scalar-tensor theories, we have to improve the dispersion relation by detuning the degeneracy conditions in an appropriate manner, where the amount of detuning should be at most of order unity in the unit of the EFT cutoff so that the Ostrogradsky ghost shows up only above the cutoff.
It is natural to expect the violation of degeneracy conditions because the degeneracy is not protected by symmetry in general.
Therefore, one needs to take into account its effect for describing stealth solutions consistently within the framework of scalar-tensor theories.
Such a controlled detuning of the imposed degeneracy conditions, dubbed ``scordatura,'' had already been taken into account in ghost condensation~\cite{ArkaniHamed:2003uy,Arkani-Hamed:2003juy,Mukohyama:2005rw} and was recently revisited in a broader context~\cite{Motohashi:2019ymr,Gorji:2020bfl,Gorji:2021isn}.

In this Letter, we show that the scordatura mechanism is generically intrinsic to the so-called U-DHOST theories~\cite{DeFelice:2018ewo,DeFelice:2021hps}. 
In U-DHOST theories, the degeneracy is partially broken in such a way that higher-derivative terms are degenerate {\it only} in the unitary gauge where the scalar field is spatially uniform (see also \cite{Gao:2014soa,Gao:2014fra,Gao:2018znj}).
Away from the unitary gauge, due to the violation of the degeneracy conditions, apparently there is an extra Ostrogradsky mode.
Nevertheless, this apparently existing Ostrogradsky mode is actually harmless as it does not propagate: It satisfies a three-dimensional elliptic differential equation on a spacelike hypersurface, meaning that the configuration of the extra mode is completely fixed by an appropriate boundary condition.
This nonpropagating mode living on a spacelike hypersurface is called a ``shadowy'' mode.
As we shall see below, the deviation from DHOST theories can be characterized by two parameters (called below ${c_{\rm D1}}$ and ${c_{\rm D2}}$), both of which are associated with the shadowy mode.
One of the parameters (${c_{\rm D1}}$) turns out to control the scordatura effect, and hence the perturbations about stealth solutions can be weakly coupled in U-DHOST theories.


{\it The model}.---We study a general class of higher-order scalar-tensor theories described by the following action:
	\begin{align}
	S[g_{\mu \nu},\phi]=\int {\rm{d}}^4x\sqrt{-g}\,\Bigg[&P(X)+Q(X)\Box \phi+F(X)R \nonumber \\
	&+\sum_{I=1}^{5}A_I(X)L_I^{(2)}\Bigg]\,, \label{HOST}
	\end{align}
where $R$ is the Ricci scalar associated with the metric~$g_{\mu \nu}$ and $\phi$ is a scalar field.
Here, $P$, $Q$, $F$, and $A_I$ ($I=1,\cdots,5$) are arbitrary functions of $X\coloneqq \phi_\mu\phi^\mu$ and 
	\begin{align}
	&L_1^{(2)}\coloneqq \phi^{\mu \nu}\phi_{\mu \nu}, \quad
	L_2^{(2)}\coloneqq (\Box\phi)^2, \quad
	L_3^{(2)}\coloneqq \phi^\mu\phi_{\mu \nu}\phi^\nu\Box\phi, \nonumber \\
	&L_4^{(2)}\coloneqq \phi^\mu\phi_{\mu \nu}\phi^{\nu\lambda}\phi_\lambda, \quad
	L_5^{(2)}\coloneqq (\phi^\mu\phi_{\mu \nu}\phi^\nu)^2,
	\end{align}
with $\phi_\mu\coloneqq {\nabla}_\mu\phi$ and $\phi_{\mu \nu}\coloneqq {\nabla}_\mu{\nabla}_\nu\phi$.
In general, the action~\eqref{HOST} yields higher-order equations of motion, and hence there exists an associated Ostrogradsky ghost~\cite{Woodard:2015zca}.
From the EFT viewpoint, the ghost is irrelevant so long as its mass scale is higher than the cutoff.
It is known that the ghost is intrinsically absent in DHOST theories, which satisfy the following degeneracy conditions~\cite{Langlois:2015cwa}:
    \begin{equation}
    {c_{\rm D1}}={c_{\rm D2}}={c_{\rm U}}=0. \label{DC}
    \end{equation}
Here, we have defined
    \begin{equation}
    \begin{split}
    {c_{\rm D1}}&\coloneqq X(A_1+A_2), \\
    {c_{\rm D2}}&\coloneqq X^2A_4+4X\Upsilon F_X-2{\mathcal{G}}_T+2(1-\Upsilon)\tilde{\mathcal{G}}_T, \\
    {c_{\rm U}}&\coloneqq -X^3A_5-X^2(A_3+A_4)-{c_{\rm D1}} \\
    &\quad~+3\Upsilon^2(2{\mathcal{G}}_T+3{c_{\rm D1}}),
    \end{split}
    \end{equation}
with a subscript $X$ denoting the $X$-derivative and
    \begin{equation}
    \begin{split}
    &{\mathcal{G}}_T\coloneqq F-XA_1, \quad
    \tilde{\mathcal{G}}_T\coloneqq (1-\Upsilon)F-2XF_X, \\
    &\Upsilon\coloneqq -\frac{X(4F_{X}+2A_2+XA_3)}{2(2{\mathcal{G}}_T+3{c_{\rm D1}})}.
    \end{split} \label{GTUp}
    \end{equation}
Specifically, the ghost mass is inversely proportional to (the square root of) ${c_{\rm U}}$ [see Eq.~\eqref{disp_rel_ghost}], meaning that the ghost is absent if only ${c_{\rm U}}=0$.
The class of theories satisfying
    \begin{align}
    {c_{\rm U}}=0 \quad\text{and}\quad
    ({c_{\rm D1}},{c_{\rm D2}})\ne (0,0)
    \label{DCU}
    \end{align}
is known as the U-DHOST theory~\cite{DeFelice:2018ewo}, in which there is no higher time derivative in the unitary gauge, but higher spatial derivatives show up, leading to the appearance of the shadowy mode~\cite{DeFelice:2018ewo,DeFelice:2021hps}.
We shall argue that the higher spatial derivatives can cure the strong coupling issue, i.e., scordatura is built-in in U-DHOST theories (except a fine-tuned subset for which $c_{\rm D1}=0$), focusing on perturbations about a stealth Minkowski solution.


{\it Stealth Minkowski perturbations}.---Let us consider general theories described by the action~\eqref{HOST} without restricting ourselves to (U-)DHOST theories from the outset.
This class of theories admits the Minkowski spacetime with a linearly time-dependent scalar field,
    \begin{equation}
    \bar{g}_{\mu \nu} {\rm{d}} x^\mu {\rm{d}} x^\nu
    =-{\rm{d}} t^2+\delta_{ij}{\rm{d}} x^i{\rm{d}} x^j, \qquad
    \bar{\phi}=qt,
    \end{equation}
as an exact solution if the function~$P(X)$ satisfies~\cite{Takahashi:2020hso}
    \begin{equation}
    P(\bar{X})=P_X(\bar{X})=0.
    \label{cond_P}
    \end{equation}
Here, $q$ is a constant and the background value of $X$ is denoted by $\bar{X}\coloneqq -q^2\,(<0)$.

In what follows, we study scalar perturbations around this stealth Minkowski solution.
Let us write down the perturbed metric and scalar field as
    \begin{equation}
    \begin{split}
    g_{\mu \nu} {\rm{d}} x^\mu {\rm{d}} x^\nu&=-(1+2\alpha){\rm{d}} t^2+2\partial_i\chi {\rm{d}} t{\rm{d}} x^i \\
    &\quad~+\left[{(1+2\zeta)\delta_{ij}+\Delta_{ij}E}\right]{\rm{d}} x^i{\rm{d}} x^j, \\
    \phi&=qt+\delta\phi,
    \end{split}
    \end{equation}
with the perturbation variables being denoted by $\alpha$, $\chi$, $\zeta$, $E$, and $\delta\phi$. 
Here, we have defined $\Delta_{ij}\coloneqq \partial_i\partial_j-(\partial^2/3)\delta_{ij}$ with $\partial^2\coloneqq \delta^{kl}\partial_k\partial_l$.
Since $\partial_t \bar\phi=q\neq 0$, we fix the gauge DOFs by setting $E=\delta\phi=0$, which is a complete gauge fixing and hence can be imposed at the Lagrangian level~\cite{Motohashi:2016prk}.
Under this gauge choice, the quadratic Lagrangian for perturbations can be written as
    \begin{align}
    {\mathcal{L}}^{(2)}=\;&{c_{\rm U}}\dot{\alpha}^2-3(2{\mathcal{G}}_T+3{c_{\rm D1}})\left({\dot{\tilde{\zeta}}^2-\frac{2}{3}\dot{\tilde{\zeta}}\partial^2\chi}\right)-2F\tilde{\zeta}\partial^2\tilde{\zeta} \nonumber \\
    &+\alpha\left({\Sigma -{c_{\rm D2}} \partial^2}\right)\alpha+6\Theta\alpha\left({\dot{\tilde{\zeta}}-\frac{1}{3}\partial^2\chi}\right) \nonumber \\
    &-4\tilde{\mathcal{G}}_T\alpha\partial^2\tilde{\zeta}-{c_{\rm D1}} (\partial^2\chi)^2,
    \label{qLag}
    \end{align}
where a dot denotes the time derivative, we have defined
    \begin{equation}
    \Sigma\coloneqq 2X^2P_{XX}, \qquad
    \Theta\coloneqq -(-X)^{3/2}Q_X,
    \end{equation}
and performed a field redefinition
    \begin{equation}
    \tilde{\zeta}\coloneqq \zeta+\Upsilon\alpha.
    \end{equation}
It is understood that all the coefficients, which are functions of $X$, are evaluated at the background~$X=\bar{X}$.
We recall that the coefficient~${c_{\rm U}}$ is absent for U-DHOST theories.
For DHOST theories, the coefficients~${c_{\rm D1}}$ and ${c_{\rm D2}}$ are also vanishing.


{\it Dispersion relation}.---In the Fourier space, the Lagrangian~\eqref{qLag} can be written in the form 
    \begin{equation}
    {\mathcal{L}}^{(2)}=\frac{1}{2}{\mathcal{K}}_{IJ}\dot{v}^I\dot{v}^J+{\mathcal{M}}_{IJ}\dot{v}^Iv^J-\frac{1}{2}{\mathcal{W}}_{IJ}v^Iv^J,
    \end{equation}
up to total derivative, where $v^I=(\alpha,\chi,\tilde{\zeta})$. 
Here, the matrices~${\mathcal{K}}$ and ${\mathcal{W}}$ are symmetric and ${\mathcal{M}}$ is antisymmetric, with ${\mathcal{M}}$ and ${\mathcal{W}}$ depending on the wavenumber~$k$.
Note that all the coefficient matrices are constant with respect to $t$ due to the symmetry of the background.
Then, the dispersion relation is given by
    \begin{equation}
    \det\left({-\omega^2{\mathcal{K}}_{IJ}-2i\omega{\mathcal{M}}_{IJ}+{\mathcal{W}}_{IJ}}\right)=0,
    \end{equation}
or written explicitly,
    \begin{equation}
    {c_{\rm U}}\omega^4+(d_0+d_1k^2)\omega^2+d_2k^2-{c_{\rm D1}} d_3k^4=0, 
    \label{disp_rel}
    \end{equation}
where
    \begin{align}
    &d_0=\Sigma+\frac{3\Theta^2}{2{\mathcal{G}}_T+3{c_{\rm D1}}}, \quad
    d_1={c_{\rm D2}}+\frac{{c_{\rm D1}}{c_{\rm U}} F}{{\mathcal{G}}_T(2{\mathcal{G}}_T+3{c_{\rm D1}})}, \nonumber \\
    &d_2=\frac{F(\Theta^2+{c_{\rm D1}}\Sigma)}{{\mathcal{G}}_T(2{\mathcal{G}}_T+3{c_{\rm D1}})}, \quad
    d_3=\frac{2\tilde{\mathcal{G}}_T^2-{c_{\rm D1}}{c_{\rm D2}} F}{{\mathcal{G}}_T(2{\mathcal{G}}_T+3{c_{\rm D1}})}.
    \end{align}

For generic higher-order scalar-tensor theories with ${c_{\rm U}}\ne 0$, there are two dynamical DOFs, one of which is the Ostrogradsky mode.
Let us now make an order estimation of the coefficients in the dispersion relation~\eqref{disp_rel}.
Writing the nonminimal coupling~$F(X)$ in the form
    \begin{equation}
    F(X)=\frac{M_{\rm Pl}^2}{2}+f(X),
    \end{equation}
with $M_{\rm Pl}$ denoting the reduced Planck mass, we assume the following scaling:
    \begin{equation}
    \begin{split}
    &X=\hat{X}M^4, \quad
    P_{XX}=\frac{\hat{P}_{\hat{X}\hat{X}}}{M^4}, \quad
    Q=\hat{Q}M, \\
    &f=\hat{f}M^2, \quad
    A_1=\frac{\hat{A}_1}{M^2}, \quad
    A_2=\frac{\hat{A}_2}{M^2}, \\
    &A_3=\frac{\hat{A}_3}{M^6}, \quad
    A_4=\frac{\hat{A}_4}{M^6}, \quad
    A_5=\frac{\hat{A}_5}{M^{10}},
    \end{split} \label{scaling}
    \end{equation}
where all the hatted quantities are dimensionless and assumed to be ${\mathcal{O}}(1)$.
Here, $M\,(\ll M_{\rm Pl})$ is a mass scale which we regard as the EFT cutoff, i.e., $\omega/M\ll 1$ and $k/M\ll 1$.
With this scaling, the leading term of each coefficient in \eqref{disp_rel} is written as
    \begin{equation}
    \begin{split}
    &{c_{\rm U}}\simeq {\hat{c}_{\rm U}} M^2, \quad
    {c_{\rm D1}}\simeq {\hat{c}_{\rm D1}} M^2, \quad
    d_0\simeq \hat{d}_0M^4, \\
    &d_1\simeq \hat{d}_1M^2, \quad
    d_2\simeq \hat{d}_2\frac{M^6}{M_{\rm Pl}^2}, \quad
    d_3\simeq 1.
    \label{disp_rel_lead}
    \end{split}
    \end{equation}
The dimensionless quantities~${\hat{c}_{\rm U}}$, ${\hat{c}_{\rm D1}}$, $\hat{d}_0$, $\hat{d}_1$, and $\hat{d}_2$ are ${\cal O}(1)$ in general and their explicit form is written as
    \begin{equation}
    \begin{split}
    &{\hat{c}_{\rm U}}=-\hat{X}(\hat{A}_1+\hat{A}_2)-\hat{X}^2(\hat{A}_3+\hat{A}_4)-\hat{X}^3\hat{A}_5, \\
    &{\hat{c}_{\rm D1}}=\hat{X}(\hat{A}_1+\hat{A}_2), \quad
    \hat{d}_0=2\hat{X}^2\hat{P}_{\hat{X}\hat{X}}, \\
    &\hat{d}_1=2\hat{X}(\hat{A}_1+\hat{A}_2)+\hat{X}^2(\hat{A}_3+\hat{A}_4), \\
    &\hat{d}_2=-\hat{X}^3\left[{\hat{Q}_{\hat{X}}^2-2\hat{P}_{\hat{X}\hat{X}}(\hat{A}_1+\hat{A}_2)}\right].
    \end{split} \label{coeff_d}
    \end{equation}
Then, the dispersion relation~\eqref{disp_rel} yields the following two branches of solution for $\omega^2$:
    \begin{align}
    \frac{\omega_1^2}{M^2}
    &\simeq -\frac{\hat{d}_2}{\hat{d}_0}\frac{M^2}{M_{\rm Pl}^2}\frac{k^2}{M^2}
    +\frac{{\hat{c}_{\rm D1}}}{\hat{d}_0}\frac{k^4}{M^4},
    \label{disp_rel_healthy} \\
    \frac{\omega_2^2}{M^2}
    &\simeq -\frac{\hat{d}_0}{{\hat{c}_{\rm U}}},
    \label{disp_rel_ghost}
    \end{align}
up to terms of higher order in $k/M$ and/or $M/M_{\rm Pl}$.
The second branch, which is absent in the limit of ${c_{\rm U}}\to 0$, corresponds to the Ostrogradsky mode.
Nevertheless, unless $|\hat{d}_0/{\hat{c}_{\rm U}}|\ll 1$, the Ostrogradsky mode does not satisfy $\omega_2/M\ll 1$ and hence is beyond the regime of validity of the EFT.
For the first (healthy) branch, the 
$k^4$~term can be ${\cal O}(1)$, which makes the perturbations weakly coupled all the way up to the cutoff scale~$M$.
This is nothing but the scordatura mechanism~\cite{Motohashi:2019ymr}.
Of course, in order to have positive $\omega_1^2$ in the regime of $M^2/M_{\rm Pl}\ll k\ll M$, the coefficient~${\hat{c}_{\rm D1}}/\hat{d}_0$ in front of $k^4$ should be positive.
Note that $\omega_1^2$ can be negative in the infrared regime~$k\lesssim M^2/M_{\rm Pl}$ if $\hat{d}_2/\hat{d}_0>0$, leading to instability.
Nevertheless, due to the $k^4$~term, the instability shows up only in the infrared regime (like the Jeans instability) and hence is harmless.
Indeed, this infrared instability and other effects put observational or experimental upper bounds on the EFT cutoff~$M$, the strongest among which is $M\lesssim 100~{\rm GeV}$~\cite{Arkani-Hamed:2005teg}. 
Note also that the scordatura effect is essentially controlled by the parameter~${c_{\rm D1}}$.
In the limit of ${c_{\rm D1}}\to 0$, the strong coupling scale becomes infinitely low, which leads to the strong coupling problem.

For U-DHOST theories in which ${c_{\rm U}}=0$, there is only one dynamical DOF, and one can obtain a single master variable as below.
When~${c_{\rm U}}=0$, the quadratic Lagrangian in the Fourier space can be written as
    \begin{align}
    {\mathcal{L}}^{(2)}
    =\;&-3(2{\mathcal{G}}_T+3{c_{\rm D1}})\left({\dot{\tilde{\zeta}}+\frac{k^2}{3}\chi-\frac{\Theta}{2{\mathcal{G}}_T+3{c_{\rm D1}}}\alpha}\right)^2 \nonumber \\
    &+2Fk^2\tilde{\zeta}^2+(d_0+{c_{\rm D2}} k^2)\alpha^2+4\tilde{\mathcal{G}}_Tk^2\alpha\tilde{\zeta} \nonumber \\
    &+\frac{2}{3}{\mathcal{G}}_Tk^4\chi^2, \label{lagFourier}
    \end{align}
from which we observe that it would be useful to employ a new variable
    \begin{equation}
    \psi\coloneqq \dot{\tilde{\zeta}}+\frac{k^2}{3}\chi-\frac{\Theta}{2{\mathcal{G}}_T+3{c_{\rm D1}}}\alpha,
    \end{equation}
instead of $\tilde{\zeta}$.
This can be done by adding the following trivial auxiliary Lagrangian to the Lagrangian~\eqref{lagFourier}:
    \begin{equation}
    3(2{\mathcal{G}}_T+3{c_{\rm D1}})\left({\dot{\tilde{\zeta}}+\frac{k^2}{3}\chi-\frac{\Theta}{2{\mathcal{G}}_T+3{c_{\rm D1}}}\alpha-\psi}\right)^2.
    \end{equation}
After integrating out the auxiliary variables, we obtain a Lagrangian written in terms of $\psi$ only, i.e.,
    \begin{align}
    {\mathcal{L}}^{(2)}&=\frac{9(2{\mathcal{G}}_T+3{c_{\rm D1}})^2}{2k^2[{\mathcal{G}}_T(2{\mathcal{G}}_T+3{c_{\rm D1}})d_3k^2-Fd_0]} \nonumber \\
    &\quad\times\left[{(d_0+{c_{\rm D2}} k^2)\dot{\psi}^2+(d_2-{c_{\rm D1}} d_3k^2)k^2\psi^2}\right]. \label{qLag2}
    \end{align}
Hence, in the regime of $M^2/M_{\rm Pl}\ll k\ll M$, the ghost-free condition is given by $\hat{d}_0>0$, where we have also assumed that $c_{\rm D2}={\hat c}_{\rm D2} M^2$, being ${\hat c}_{\rm D2}$ of order unity.
There is an apparent ghost for smaller $k$, but it is harmless~\cite{Gumrukcuoglu:2016jbh}.
From the Lagrangian~\eqref{qLag2}, we obtain the dispersion relation as
    \begin{equation}
    (d_0+{c_{\rm D2}} k^2)\omega^2+d_2k^2-{c_{\rm D1}} d_3k^4=0,
    \end{equation}
which is nothing but the ${c_{\rm U}}\to 0$~limit of \eqref{disp_rel}.
By solving the dispersion relation for $\omega^2$, 
we have
    \begin{equation}
    \frac{\omega^2}{M^2}
    \simeq -\frac{\hat{d}_2}{\hat{d}_0}\frac{M^2}{M_{\rm Pl}^2}\frac{k^2}{M^2}
    +\frac{{\hat{c}_{\rm D1}}}{\hat{d}_0}\frac{k^4}{M^4}.
    \label{disp_rel_U-DHOST}
    \end{equation}
This essentially reproduces \eqref{disp_rel_healthy}, i.e., the dispersion relation for the healthy branch in the case of ${c_{\rm U}}\ne 0$, and hence the scordatura mechanism works so long as ${\hat{c}_{\rm D1}}$ is positive and ${\cal O}(1)$.
One may notice that the sign of the $k^2$~term is negative, leading to instability (note that $\hat{d}_2=-\hat{X}^3\hat{Q}_{\hat{X}}^2+{\hat{c}_{\rm D1}}\hat{d}_0>0$).
Nevertheless, as mentioned earlier, this instability is an infrared one and hence is harmless.
On the other hand, the sign of $\hat{c}_{\rm D2}$ will never trigger any instability unless $|\hat{c}_{\rm D2}|\gg\hat{d}_0$.
In the case of DHOST theories where ${c_{\rm D1}}={c_{\rm D2}}=0$ on top of ${c_{\rm U}}=0$, the dispersion relation reads
    \begin{equation}
    \omega^2\simeq -\frac{\hat{d}_2}{\hat{d}_0}\frac{M^2}{M_{\rm Pl}^2}k^2,
    \end{equation}
which means that the sound speed is suppressed by $M^2/M_{\rm Pl}^2$, and hence the perturbations would be strongly coupled.
This is consistent with the result of \cite{Motohashi:2019ymr}.

Our result is summarized in the \hyperref[fig]{Figure} below.
The shaded region corresponds to $|{\hat{c}_{\rm D1}}|\ll 1$ or $|{c_{\rm D1}}|\ll M^2$, where the strong coupling problem shows up.
DHOST theories have ${c_{\rm D1}}=0$ and lie within the shaded region.
On the other hand, outside the shaded region, perturbations would be weakly coupled.
U-DHOST theories have ${\hat{c}_{\rm D1}}={\mathcal{O}}(1)$ in general, and hence the scordatura mechanism is intrinsic to this class of theories.

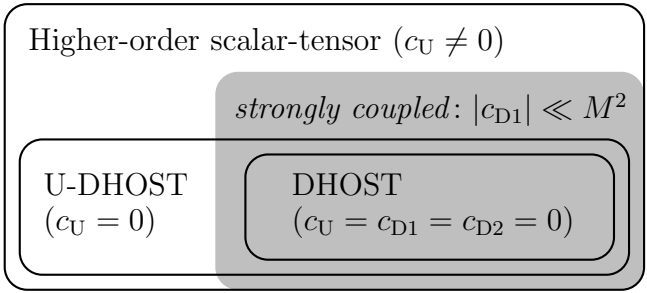
\begin{figure}[H]
\centering
\begin{tikzpicture}
	\fill[fill=lightgray, rounded corners=0.3cm] (2.8,0)--(2.8,2.9)--(8.5,2.9)--(8.5,0)--cycle;
	\node[anchor=east] at (8.4,2.4) {\large{{\it strongly coupled}\,:~$|{c_{\rm D1}}|\ll M^2$}};
	\draw[thick, rounded corners=0.3cm] (0,0)--(0,3.8)--(8.5,3.8)--(8.5,0)--cycle;
	\node[anchor=west] at (0.2,3.3) {\large{Higher-order scalar-tensor (${c_{\rm U}}\ne 0$)}};
	\draw[thick, rounded corners=0.3cm] (0.2,0.2)--(0.2,2)--(8.3,2)--(8.3,0.2)--cycle;
	\node[anchor=west] at (0.4,1.4) {\large{U-DHOST}};
	\node[anchor=west] at (0.4,0.9) {\large{(${c_{\rm U}}=0$)}};
	\draw[thick, rounded corners=0.3cm] (3.2,0.4)--(3.2,1.8)--(8.1,1.8)--(8.1,0.4)--cycle;
	\node[anchor=west] at (3.7,1.4) {\large{DHOST}};
	\node[anchor=west] at (3.7,0.9) {\large{(${c_{\rm U}}={c_{\rm D1}}={c_{\rm D2}}=0$)}};
\end{tikzpicture}
\renewcommand{\figurename}{FIG}
\renewcommand{\thefigure}{\!\!}
\vspace{-5mm}
\caption{Higher-order scalar-tensor theories and the strong coupling issue on the stealth Minkowski background.}
\label{fig}
\end{figure}
\vspace{-2mm}


{\it Shadowy mode}.---Finally, let us make a comment on the shadowy mode, which is present in the case of U-DHOST theories.
As was demonstrated in \cite{DeFelice:2021hps}, a Hamiltonian analysis is useful to see how the shadowy mode shows up.
For this purpose, we introduce the Arnowitt-Deser-Misner variables as $g_{\mu \nu} {\rm{d}} x^\mu {\rm{d}} x^\nu=-N^2{\rm{d}} t^2+\gamma_{ij}({\rm{d}} x^i+N^i{\rm{d}} t)({\rm{d}} x^j+N^j{\rm{d}} t)$,
where $N$ is the lapse function, $N^i$ is the shift vector, and $\gamma_{ij}$ is the induced metric.
Here, we take the unitary gauge where $\phi=\phi(t)$.
In this gauge, the kinetic term of the scalar field is related to the lapse function through $X=-\dot{\phi}^2/N^2$.
For simplicity, we focus on U-DHOST theories with $\Upsilon=0$.
In other words, we move to a frame where $\Upsilon=0$ by use of an invertible disformal transformation~\cite{BenAchour:2016fzp,Domenech:2015tca,Takahashi:2017zgr}.
Then, the Lagrangian density in \eqref{HOST} can be rewritten as
    \begin{align}
    {\mathcal{L}}=N\sqrt{\gamma}\,\bigg[
    &{\mathcal{G}}_TK_{ij}K^{ij}-({\mathcal{G}}_T+{c_{\rm D1}})K^2 \nonumber \\
    &+{c_{\rm D2}}\frac{{\rm D}_iN{\rm D}^iN}{N^2} +\tilde{Q}K+F{\mathcal{R}}+P\bigg]\,,
    \label{lagADM}
    \end{align}
where ${\mathcal{R}}$ is the spatial curvature, $K_{ij}$ is the extrinsic curvature with its trace denoted by $K\coloneqq K^i{}_i$, and ${\rm D}_i$ is the covariant derivative associated with $\gamma_{ij}$.
Here, the function~$\tilde{Q}(X)$ is defined so that $\tilde{Q}_X=-(-X)^{-1/2}Q_X$.
Performing a Legendre transformation, the total Hamiltonian can be straightforwardly obtained as
    \begin{equation}
    H_T=\int {\rm{d}}^3x\left({{\mathcal{H}}_N+N^i{\mathcal{H}}_i+u_N\pi_N+u^i\pi_i}\right),
    \end{equation}
where the canonical momenta conjugate to $N$, $N^i$, $\gamma_{ij}$ are, respectively, denoted by $\pi_N$, $\pi_i$, $\pi^{ij}$, and we have defined
    \begin{align}
    {\mathcal{H}}_N&\coloneqq N\sqrt{\gamma}\,\bigg[\frac{1}{{\mathcal{G}}_T}\left({\frac{\pi_{ij}\pi^{ij}}{\gamma}-\frac{{\mathcal{G}}_T+{c_{\rm D1}}}{2{\mathcal{G}}_T+3{c_{\rm D1}}}\frac{\pi^2}{\gamma}}\right) \nonumber \\
    &\qquad\qquad~~ +\frac{\tilde{Q}}{2{\mathcal{G}}_T+3{c_{\rm D1}}}\left({\frac{\pi}{\sqrt{\gamma}}-\frac{3}{4}\tilde{Q}}\right) \nonumber \\
    &\qquad\qquad~~ -{c_{\rm D2}}\frac{{\rm D}_iN{\rm D}^iN}{N^2}
    -F{\mathcal{R}}-P\bigg]\,, \label{hamC} \\
    {\mathcal{H}}_i&\coloneqq -2{\rm D}^j\frac{\pi_{ij}}{\sqrt{\gamma}},
    \end{align}
with $\pi\coloneqq \pi^i{}_i$.
The primary constraints~$\pi_N\approx 0$ and $\pi_i\approx 0$ have been incorporated in the total Hamiltonian by use of the Lagrangian multipliers~$u_N$ and $u^i$, respectively.
Although the momentum constraints~${\mathcal{H}}_i\approx 0$ remain first class, the Hamiltonian constraint is now second class, which fixes $N$ as a functional of other canonical variables.
As is clearly seen in \eqref{hamC}, a nonvanishing ${c_{\rm D2}}$ results in higher spatial derivatives in the Hamiltonian constraint, which reflects the presence of the shadowy mode (see also \cite{DeFelice:2021hps}).
Note also that the terms quadratic in $\pi_{ij}$ are detuned from the usual combination~$\pi_{ij}\pi^{ij}-\pi^2/2$ when ${c_{\rm D1}}\ne 0$, as in nonprojectable Ho\v{r}ava-Lifshitz gravity~\cite{Horava:2009uw}.
Therefore, even for the fine-tuned case~${c_{\rm D2}}=0$, there would be a shadowy mode corresponding to the instantaneous mode in nonprojectable Ho\v{r}ava-Lifshitz gravity~\cite{Blas:2010hb,Blas:2011ni}.


{\it Discussions}.---In this Letter, we studied the perturbations about the stealth Minkowski solution in higher-order scalar-tensor theories described by the action~\eqref{HOST}.
We found that the perturbations would be strongly coupled in DHOST theories satisfying the degeneracy conditions~${c_{\rm D1}}={c_{\rm D2}}={c_{\rm U}}=0$, while this problem can be avoided in nondegenerate higher-order scalar-tensor theories in general, which is consistent with the result of \cite{Motohashi:2019ymr}.
We clarified that the parameter~${c_{\rm D1}}$ controls the magnitude of the sound speed and thereby the strong coupling scale.
Our result implies that the scordatura effect is intrinsic to generic U-DHOST theories, for which ${c_{\rm U}}=0$ and ${c_{\rm D1}}={\mathcal{O}}(M^2)$.
Also, the dispersion relation for generic higher-order scalar-tensor theories is the same as the one for U-DHOST theories at the leading order of $M/M_{\rm Pl}$.
Moreover, the dispersion relation with higher-$k$ operators obtained for the Minkowski background would apply to any short-wavelength modes in any curved backgrounds.
Indeed, in a curved background, the dispersion relation would be subject to a correction of the form
    \begin{equation}
    \delta\omega^2=\Lambda\left[a_0+a_2\left(\frac{k}{M}\right)^2+a_4\left(\frac{k}{M}\right)^4+\cdots\right],
    \end{equation}
where $\Lambda$ represents a typical scale of the curvature (e.g., the effective cosmological constant in the case of stealth de Sitter solutions) and $a_i$'s are dimensionless constants of ${\mathcal{O}}(1)$.
Note that we assume $\Lambda/M^2\ll 1$ for the EFT to be valid.
Compared to the $k^4$~term (i.e., the scordatura term) in the dispersion relation~\eqref{disp_rel_U-DHOST} for the Minkowski case, one can ignore the effect of curved spacetime so long as $M^2\Lambda\ll k^4\,(\ll M^4)$.
Hence, U-DHOST theories would provide the most general effective framework that describes perturbations about stealth background.
It would be intriguing to further study stealth solutions, e.g., black holes along the lines of \cite{deRham:2019gha,Khoury:2020aya,Takahashi:2021bml} within the framework of U-DHOST theories.


The work of A.D.F.~was supported by the Japan Society for the Promotion of Science~(JSPS) Grants-in-Aid for Scientific Research No.~20K03969. 
S.M.'s work was supported in part by JSPS Grants-in-Aid for Scientific Research No.~17H02890, No.~17H06359, and by World Premier International Research Center Initiative, MEXT, Japan. 
The work of K.T.~was supported by JSPS KAKENHI Grant No.~JP21J00695.


\bibliographystyle{mybibstyle}
\bibliography{bib}

\end{document}